# The lattice and electronic thermal conductivity of doped SnSe: a first-principles study

Shouhang Li, Zhen Tong and Hua Bao[*]

University of Michigan-Shanghai Jiao Tong University Joint Institute, Shanghai Jiao Tong University, Shanghai 200240, P. R. China

Recently, it has been found that crystalline tin selenide (SnSe) holds great potential as a thermoelectric material due to its ultralow thermal conductivity and moderate electronic transport performance. As thermoelectric application usually requires doped material, charge carriers can play a role in the thermal transport in doped SnSe, but such an effect has not been clearly elucidated in previous theoretical and experimental studies. Here we performed a fully first-principles study on the effects of electrons to the thermal transport in doped SnSe. The electron-phonon coupling (EPC) effects on both phonons and charge carriers were considered using the mode specific calculation in our work. It is found that for phonons, EPC are weak compared to the intrinsic phonon-phonon scattering even at high carrier concentrations and thus have negligible effects on the lattice thermal conductivity. The electronic thermal conductivity is not negligible when the carrier concentration is higher than $10^{19}$ cm$^{-3}$ and the values can be as high as 1.55, 1.45 and 1.77 Wm$^{-1}$K$^{-1}$ on *a*, *b* and *c* axes, respectively, for $10^{20}$ cm$^{-3}$ electron concentration at 300K. The Lorenz number of SnSe is also calculated and it is dependent on crystal orientations, carrier concentrations, and carrier types. The simple estimation of electronic thermal conductivity using Wiedemann-Franz law can cause large uncertainties for doped SnSe.

---

[*] To whom correspondence should be addressed. Email: hua.bao@sjtu.edu.cn (HB)

The harvest of low grade waste heat is of great importance to the sustainable usage of energy. Thermoelectric (TE) materials are placed great expectations on achieving this goal [1]. Recently, SnSe has attracted enormous research interest due to its excellent TE performance [2, 3]. In 2014, Zhao *et al.* found that intrinsic SnSe crystal holds extremely low thermal conductivity which leads to its impressively high figure of merit (ZT) value of ~2.6 at 923K [2]. Later, it is verified that the heavily doped *p*-type SnSe owns very high ZT value along the *b*-axis due to its excellent electronic transport properties [4]. The out-of-plane direction for bromine doped *n*-type SnSe has even better ZT performance, which reaches ~2.8 at 773K due to its ultralow thermal conductivity, high mobility and large Seebeck coefficient [3]. To explain the ultralow lattice thermal conductivity, the giant anharmonicity of this material has been suggested by theoretical studies [5, 6] and verified by experiments [7, 8].

The reported high ZT values are usually achieved with heavily doped SnSe and the doping concentration is usually larger than $10^{19}$ cm$^{-3}$ [2, 4, 6]. The electron effects on the thermal conductivity cannot be simply neglected under such condition. On one hand, the charge carriers can scatter phonons and thus further reduce the lattice thermal conductivity. For example, Liao *et al.* [9] found that the lattice thermal conductivity of silicon is significantly reduced by charge carrier scatterings when the carrier concentration goes over $10^{19}$ cm$^{-3}$. On the other hand, it is possible that the electronic thermal conductivity can play a role when the carrier concentration is high. In experiment, Zhao *et al.* [4] found the electronic thermal conductivity component of 1.5 mol% sodium doped SnSe reaches ~0.94 Wm$^{-1}$K$^{-1}$ on the *b*-axis at the temperature of 300K, which is larger than the lattice thermal conductivity of ~0.70 Wm$^{-1}$K$^{-1}$. To estimate the carrier effects on thermal conductivity of SnSe, the Wiedemann-Franz (WF) law [2, 6, 10, 11] and constant relaxation time approximation (CRTA) [12, 13] were used, and it is roughly suggested that the electron thermal conductivity component cannot be neglected at a carrier concentration larger than $3\times10^{19}$ cm$^{-3}$ [6]. However, such methods have been verified to hold large deviation for SnSe [14] and other thermoelectric materials [15-17]. In order to better understand the thermal transport mechanism in doped SnSe, it is necessary to perform a detailed mode specific analysis to investigate the effects of charge carriers on the thermal conductivity.

In this letter, we report the mode-by-mode analysis of both lattice and electronic thermal conductivity of doped SnSe, employing the first-principles calculations with a full consideration of phonon-phonon and electron-phonon scatterings. In the lattice thermal conductivity part, we considered both phonon-phonon (PPI) and electron-phonon (EPI) interactions. Then, in the electronic thermal conductivity part, we consider charge carrier scattering by phonons. The calculated lattice and electronic thermal conductivity for both *n*-type and *p*-type SnSe in the carrier concentration range of $10^{17}$-$10^{20}$ cm$^{-3}$ are presented. The results are compared and discussed with previous theoretical and experimental works.

By combining the phonon BTE and the Fourier's law [18-20], the lattice thermal conductivity can be expressed as

$$\kappa_{ph,\alpha\beta} = \sum_\lambda c_{v,\lambda} v_{\lambda,\alpha} v_{\lambda,\beta} \tau_\lambda = \frac{1}{V} \sum_\lambda \hbar\omega_\lambda \frac{\partial n_\lambda}{\partial T} v_{\lambda,\alpha} v_{\lambda,\beta} \tau_\lambda, \quad (1)$$

where $\lambda = (\mathbf{q}, \nu)$ denotes the phonon mode with wave vector $\mathbf{q}$ and polarization $\nu$, $c_{v,\lambda}$ is the volumetric specific heat, $v_{\lambda,\alpha}$ is the $\alpha$ component of the phonon group velocity, $\tau_\lambda$ is the phonon lifetime, $V$ is the volume of the unit cell, $\hbar$ is the reduced Planck's constant, $\omega_\lambda$ is the phonon frequency, $T$ is the temperature, and $n_\lambda$ is the Bose-Einstein distribution.

The crucial step is to calculate the phonon relaxation time $\tau_\lambda$, which is related to several scattering processes, here we only consider PPI and EPI. The total phonon relaxation time can be obtained by using Matthiessen's rule [21] as $1/\tau_\lambda = 1/\tau_\lambda^{PPI} + 1/\tau_\lambda^{EPI}$. $1/\tau_\lambda^{PPI}$ denotes the phonon-phonon scattering rate which is related to the three-phonon scattering matrix element [22]. $1/\tau_\lambda^{EPI}$ is the electron-phonon scattering rate which is related to the electron-phonon scattering matrix element [23]. It is determined by $1/\tau_\lambda^{EPI} = 2\mathrm{Im}(\Pi)/\hbar$. $\mathrm{Im}(\Pi)$ denotes the imaginary part of the phonon self-energy $\Pi$, which can be expressed as

$$\Pi_\lambda(\omega, T) = 2 \sum_{mn} \int \frac{d\mathbf{k}}{\Omega} |g_{mn,\nu}(\mathbf{k},\mathbf{q})|^2 \times \frac{f_{n\mathbf{k}}(T) - f_{m\mathbf{k}+\mathbf{q}}(T)}{\omega - (\varepsilon_{m\mathbf{k}+\mathbf{q}} - \varepsilon_F) - \omega_\lambda + i\delta}. \quad (2)$$

The matrix element $g_{mn,\nu}(\mathbf{k}, \mathbf{q})$ quantifies the scattering processes between the electronic state $n\mathbf{k}$ and $m\mathbf{k+q}$, $\Omega$ is the volume of the first Brillouin zone, $f_{n\mathbf{k}}$ and $f_{m\mathbf{k+q}}$ stand for Fermi-Dirac distribution, $\varepsilon_{m\mathbf{k+q}}$ is the electron energy, $\varepsilon_F$ is the Fermi energy, $\delta$ is a small positive parameter introduced to guarantee the numerical stability.

The electronic thermal conductivity can be extracted from the Onsager relations [24, 25]. Here we obtained electron relaxation time by considering electron-phonon coupling [23]. The electron relaxation time is determined by $1/\tau_{m\mathbf{k}}^{ep} = 2\mathrm{Im}(\Sigma)\hbar$. $\mathrm{Im}(\Sigma)$ denotes the imaginary part of the electron self-energy $\Sigma$, which can be expressed as

$$\Sigma_{m\mathbf{k}}(\omega,T) = \sum_{nv}\frac{d\mathbf{q}}{\Omega}\left|g_{mn,v}(\mathbf{k},\mathbf{q})\right|^2 \times \left[\frac{n_\lambda(T)+f_{m\mathbf{k+q}}(T)}{\omega-(\varepsilon_{m\mathbf{k+q}}-\varepsilon_F)+\omega_\lambda+i\delta} + \frac{n_\lambda(T)+1-f_{m\mathbf{k+q}}(T)}{\omega-(\varepsilon_{m\mathbf{k+q}}-\varepsilon_F)-\omega_\lambda+i\delta}\right]. \quad (3)$$

The symbols have the same meaning as in Eqs. (1) and (2).

All our first-principles calculations are carried out with Quantum Espresso [26]. A Perdew-Burke-Ernzerhof (PBE) form [27] of generalized gradient approximation (GGA) was employed as the exchange-correlation functional. The cutoff energy of the plane wave was set as 45 Ry, and a $6 \times 10 \times 10$ Monkhorst-Pack (MP) k-points mesh was used for the self-consistent filed calculation. The convergence threshold of energy was set to be $10^{-12}$ Ry.

We followed the workflow proposed by Li *et al.* [22] to calculate the phonon thermal conductivity. For harmonic force constant, the density-functional perturbation theory was used [28]. The **q**-points mesh was set as $3 \times 3 \times 3$, the energy threshold for phonon calculation was set to be $10^{-16}$ Ry to guarantee the convergence. For the cubic force constant, it was extracted through the combination of density functional theory calculation and the finite displacement method, as demonstrated in Ref. [22]. A supercell of $2 \times 3 \times 3$ was used and the large 18th nearest neighbors were included for the third-order interactions to cover the long-ranged dipole-dipole interactions, which was corresponding to a cutoff distance of ~6.35Å. For the electronic properties calculations, the electron-phonon scattering rate was calculated by the Electron-Phonon Wannier (EPW) package [23]. It is based on the maximally localized Wannier functions (MLWFs) and generalized Fourier interpolation [29]. The electron band structure, the phonon spectrum and phonon potential differential as well as the electron-phonon coupling matrix elements are first calculated on the coarse grids of $6 \times 6 \times 6$ **k**-points and $3 \times 3 \times 3$ **q**-points, and are then interpolated to the dense $40 \times 40 \times 40$ **k**-points and $25 \times 25 \times 25$ **q**-points.

SnSe holds *Pnma* space group at room temperature. The orthorhombic lattice structure is shown in the inset of Fig. 1. After the lattice structure optimization, we obtain the unit cell lengths *a*=11.76Å, *b*=4.22Å, *c*=4.52Å, which are consistent with the results reported by others using a similar method [5-7]. It is well-known that density functional theory calculations underestimate the bandgap for semiconductors [30], and it would have a strong effect on the electronic transport properties. The rigid band approximation

(RBA) [31] was applied here to shift the entire conduction band uniformly to open up the bandgap and match the experimentally measured value of 0.86 eV [2]. With RBA, the carrier concentration is related to the Fermi energy [32] and it is equal to the integration of the production between electron density of state and Fermi-Dirac distribution. The electron and hole concentrations are later denoted as $n_e$ and $n_h$, respectively.

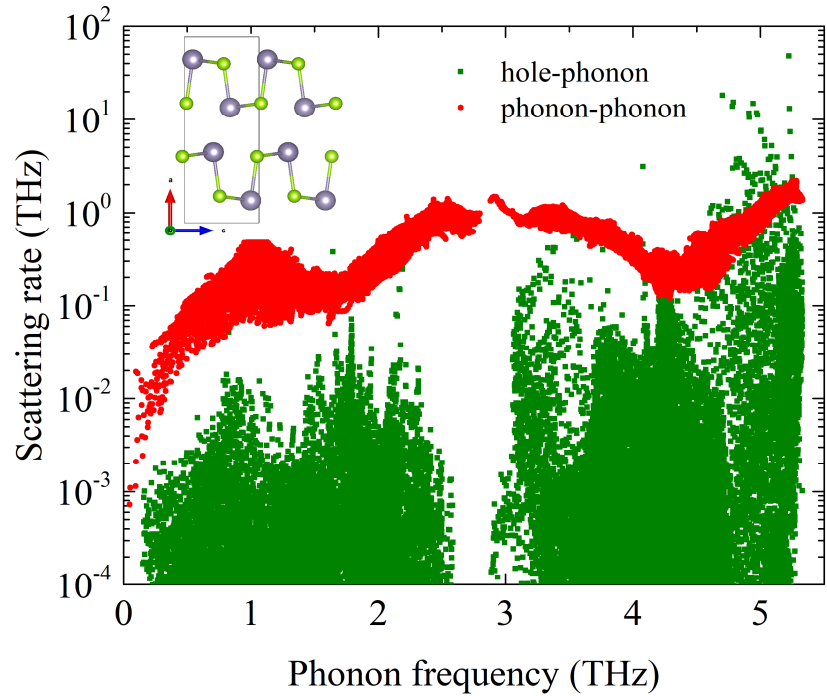

Fig. 1 The phonon scattering rate $1/\tau_\lambda^{PPI}$ (red) and $1/\tau_\lambda^{EPI}$ (green) vary with phonon frequency. We take $1/\tau_\lambda^{EPI}$ at the hole concentration of $10^{20}$ cm$^{-3}$ as an example. The inset of this figure is the lattice structure of orthorhombic lattice structure (*Pnma* phase) of SnSe. The color gray stands for Sn atoms and green stands for Se atoms. The box indicates a conventional unit cell of SnSe, which contains 4 Sn atoms and 4 Se atoms.

By only considering PPI (intrinsic SnSe), we found that the lattice thermal conductivity at 300K is 0.72, 1.84 and 1.39 Wm$^{-1}$K$^{-1}$ on the *a*, *b*, and *c* axes, respectively. The thermal conductivity on the *a* axis is much smaller than that on the *b* and *c* axes. The ultralow thermal conductivity is related to the strong anharmonic lattice structure of SnSe. It can be also supported by the relatively large Grüneisen parameter $\gamma$, which is determined to be 1.49 by our calculation. The predicted lattice thermal conductivity values are larger than those reported by Zhao *et al.* [2], which are 0.46, 0.70 and 0.68 Wm$^{-1}$K$^{-1}$, but our results are consistent with the previous theoretical studies [5, 6] and some other recent experimental reports [3, 10]. The discussions on the extremely low lattice thermal conductivity of SnSe is very convincing in the Refs. [5, 6, 33], so they are not repeatedly discussed here. In below, we focus on the electron effects on the thermal conductivity.

Considering the phonon-electron process, we determined that the lattice thermal conductivity is reduced to 0.72, 1.80, and 1.36 Wm$^{-1}$K$^{-1}$ at 300K for the $10^{20}$ cm$^{-3}$ electron carrier case on the *a*, *b*, and *c* axes, respectively, while the values are 0.71, 1.82 and 1.38 Wm$^{-1}$K$^{-1}$ for the hole carrier case. These values are only slightly smaller than the case when we only consider PPI. This indicates that EPI is much weaker than PPI. This is further supported by Fig. 2, where one can see that carrier concentration has little effects on the lattice thermal conductivity even when the carrier concentration is higher than $10^{19}$ cm$^{-3}$. The scattering rate related to PPI and EPI are shown in Fig. 1. One can see the electron-phonon scattering rate $1/\tau_\lambda^{EPI}$ due to EPI is much smaller than the phonon-phonon scattering rate $1/\tau_\lambda^{PPI}$. They are on an average value of $5.53 \times 10^{-3}$ THz and 0.43 THz, respectively. This is different from the strong Fröhlich effects on wurtzite GaN [34], Si [9], and metals [35, 36]. SnSe is different from Si and GaN because it has stronger lattice anharmonicity, which gives quite large $1/\tau_\lambda^{PPI}$, while the magnitude of $1/\tau_\lambda^{EPI}$ of SnSe is smaller than that of Si [9]. It is also different from metals because the concentration of carriers in doped semiconductors is still much smaller as compared to metals.

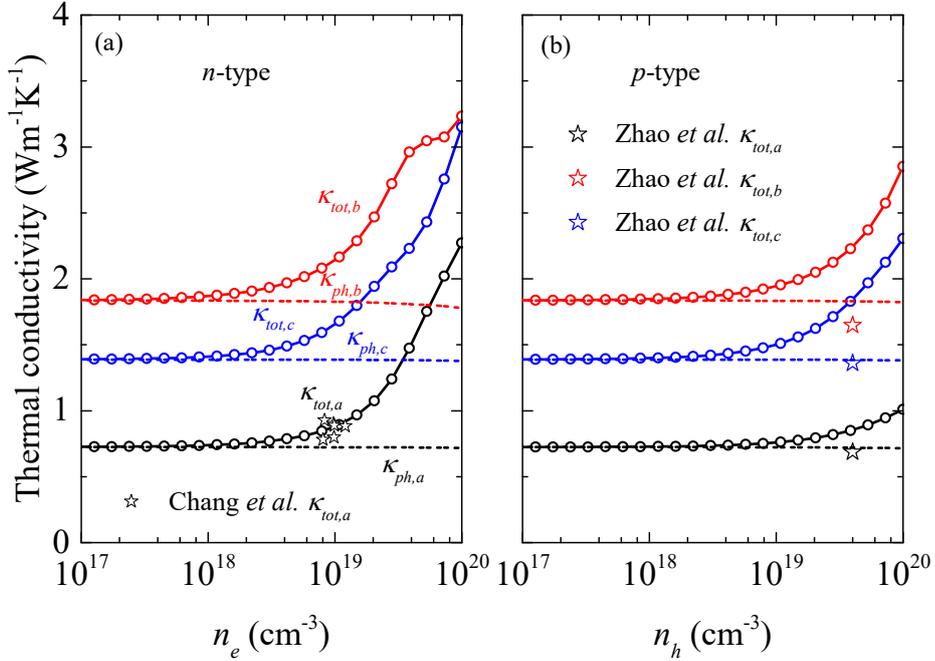

Fig. 2 The lattice thermal conductivity (dash lines) and total thermal conductivity (solid lines) along the *a*, *b*, and *c* axes with different carrier concentrations at 300K for (a) *n*-type and (b) *p*-type SnSe. $\kappa_{tot,\alpha}$ and $\kappa_{ph,\alpha}$ denote the total and phonon thermal conductivity on the $\alpha$ crystal orientation, respectively. The stars in (a) and (b) are experimental results, taken from Ref. [3] and Ref. [4], respectively.

The electronic thermal conductivity is also predicted based on the mode specific electron relaxation time and electron BTE, and the results are shown in Fig. 3. We can see the electronic thermal conductivity increases quickly with the carrier concentration on the three axes. By adding the lattice and electronic thermal conductivity components together, we found that the electronic contribution becomes important when the carrier concentration is larger than $10^{19}$ cm$^{-3}$. This can also be more clearly seen in Fig. 2. The electronic thermal conductivities achieve 1.55, 1.45, and 1.77 Wm$^{-1}$K$^{-1}$ on the *a*, *b* and *c* axes which is comparable to the phonon thermal conductivity for *n*-type SnSe at the concentration of $10^{20}$ cm$^{-3}$. Similar observation is also observed for *p*-type. The anisotropy of the electronic thermal conductivity is smaller for the *n*-type SnSe than *p*-type. Especially, for *p*-type SnSe, the thermal conductivity on the *a*-axis is significantly smaller than that on the *b* and *c* axes. This should be related to the large difference between the electron conduction bands and valence bands [37]. Our predicted electronic thermal conductivity for the *n*-type is slightly larger than the experimental results reported by Ref. [3]. The predicted total thermal conductivity value is very close to their results on the *a* axis, as shown in Fig. 2 (a). For the *p*-type, our predicted value is smaller than the experimental results in Ref. [4]. However, the total thermal conductivity is still larger than theirs due to their extremely low lattice thermal conductivity, as shown in Fig 2 (b).

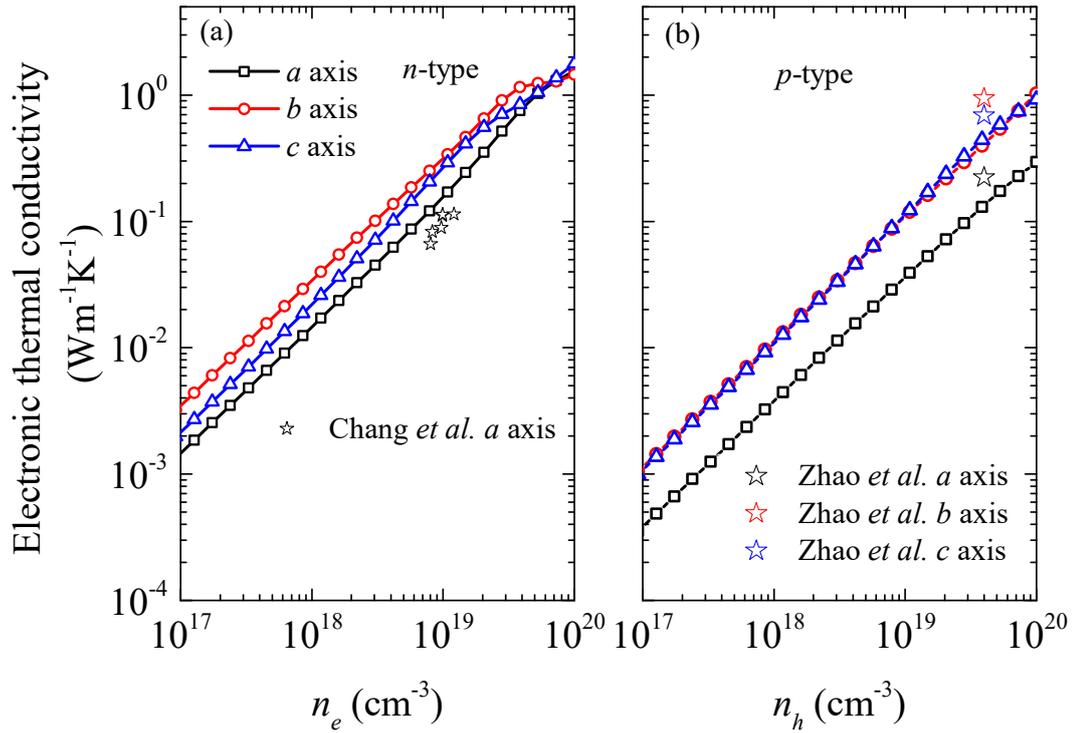

Fig. 3 The electronic thermal conductivity component along the *a*, *b*, and *c* axes with different carrier concentrations at 300K for (a) *n*-type and (b) *p*-type. The pentagram scatters in (a) and (b) are experimental results, taken from Ref. [3] and Ref. [4], respectively.

We also predicted the electrical conductivity of both *n*- and *p*-type SnSe, as shown in Fig. 4. The electrical conductivity reaches 1843, 2620 and 2071 S/cm for the *n*-type at the concentration of $10^{20}$ cm$^{-3}$, while the values are 513, 1672 and 1550 S/cm for the *p*-type. The anisotropy is also observed in the electrical conductivity. The electrical conductivity on the *a* axis is significantly smaller than that on the *b* and *c* axes at the low and mediate carrier concentration for both *n*-type and *p*-type. However, as shown in Fig. 4 (a), the electrical conductivity on the *a* axis is dramatically larger for the *n*-type when the carrier concentration is close to $5 \times 10^{19}$ cm$^{-3}$. The tendency was also reported in Ref. [37]. Besides, the electrical conductivity on the *a* axis for *n*-type is significantly larger than that on the *p*-type, which suggests the good electronic transport performance for *n*-type and it holds great potential for thermoelectric application [3]. This can also interpret the relatively larger electronic thermal conductivity on *a* axis for *n*-type compared to that of *p*-type shown in Fig. 3. The values on the *b* and *c* axes are also larger for *n*-type at the same concentration. This is consistent with a recent report [14]. For comparison, the electrical conductivity predicted by different theoretical studies including CRTA [13], single parabolic band (SPB) model [6], and measured by experiments [3, 4] are presented in Fig. 4. The CRTA method significantly underestimates the electrical conductivity for *n*-type SnSe, as shown in Fig. 4 (b). SPB model is close to our mode specific method more or less. However, SPB model has larger derivation for both *n*- and *p*-type SnSe. Our predicted results of the electrical conductivity for the *n*-type SnSe on the *a* axis is very close to the experimental results reported by Ref. [3]. Ref. [4] reported larger electrical conductivity for *p*-type SnSe compared to our mode specific method and SPB method [6], while our results are closer to the experimental results compared to the SPB method. This indicates that there exist some uncertainties in both CRTA and SPB models.

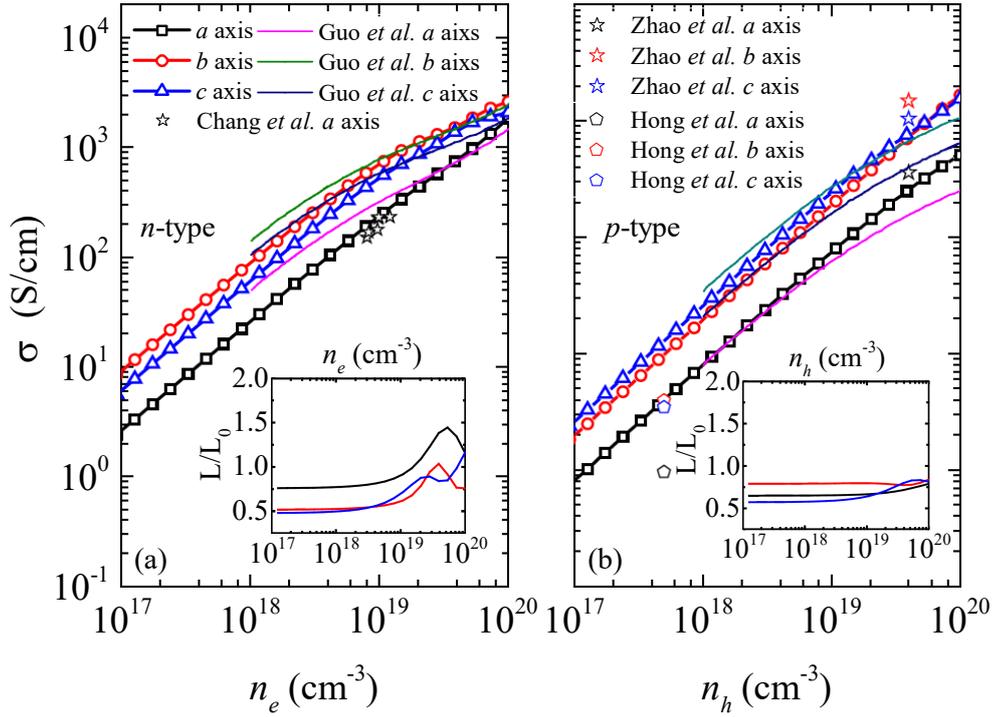

Fig. 4 The electrical conductivity along the *a*, *b*, and *c* axes with different carrier concentrations at 300K for (a) *n*-type and (b) *p*-type. The thin solid line data in (a) and (b) are estimated using SPB model, taken from Ref. [6]. The pentagon scatters in (b) are predicted using constant relaxation time approximation, taken from Ref. [13]. The stars scatters in (a) and (b) are experimental results taken from Ref. [3] and Ref. [4], respectively. The inset figures in (a) and (b) are the normalized Lorenz numbers ($L/L_0$) for *n*- type and *p*-type, respectively, where $L_0 = 2.44 \times 10^{-8} W\Omega K^{-2}$.

The Wiedemann-Franz law is commonly used to separate the electronic and lattice thermal conductivity components, i.e. $\kappa_{ph} = \kappa_{tot} - L\sigma T$. However, The Lorenz number $L$ is in a wide range for different thermoelectric materials [17], which can have large deviation to the widely adopted value $L_0 = 2.44 \times 10^{-8} W\Omega K^{-2}$. For SnSe at ~300K, Chang *et al.* [3] used $L$ as $1.61 \sim 1.64 \times 10^{-8} W\Omega K^{-2}$, Zhao *et al.* [4] used $L$ as $2.00 \times 10^{-8} W\Omega K^{-2}$. The value $2.45 \times 10^{-8} W\Omega K^{-2}$ was used by Ibrahim *et al.*

[10] to evaluate the electronic thermal conductivity for SnSe at 800K. The wide range of the Lorenz number $L$ can lead to large differences in predicting electronic thermal conductivity. Based on our calculation results, it is found that $L$ is strongly related to crystal orientations, carrier concentrations, and carrier types, as shown in the insets of Fig. 4. The Lorenz number can be regarded as a constant which is only related to the crystal orientations and carrier types when the carrier concentration is smaller than $5 \times 10^{18}$ cm$^{-3}$. However, it holds a complex behavior in the heavy doping concentration range, in which people who work on thermoelectric properties mainly focus on. It suggests that our more precise mode specific method is valuable to detect the accurate electronic thermal transport properties of SnSe.

In summary, we performed a first-principles study to predict the lattice and electronic thermal conductivity of doped SnSe. The effects of electron phonon coupling on thermal conductivity are carefully discussed. For the lattice thermal conductivity component, it is found that SnSe holds the extremely low value of 0.72, 1.80 and 1.36 Wm$^{-1}$K$^{-1}$ on the $a$, $b$, and $c$ axes, respectively, for $10^{20}$ cm$^{-3}$ electron concentration at 300K. The electron-phonon coupling has little effect on lattice thermal conductivity, which is attributed to the strong anharmonic phonon-phonon interaction. The electronic thermal conductivity is not negligible when the concentration is larger than $10^{19}$ cm$^{-3}$ for both $n$-type and $p$-type. The values reach 1.55, 1.45 and 1.77 Wm$^{-1}$K$^{-1}$ on the $a$, $b$, and $c$ axes, respectively, for the $10^{20}$ cm$^{-3}$ electron concentration case, which is comparable to the lattice thermal conductivity. Finally, we showed that the Lorenz number is crystal orientations, carrier concentrations, and carrier types dependent. The using a constant Lorenz number can introduce large uncertainties in estimating the electronic thermal conductivity of SnSe.

We would like to thank Dr. Dengdong Fan and Dr. Xiaokun Gu for valuable discussions. This work was supported by the National Natural Science Foundation of China No. 51676121 (HB). Simulations were performed with computing resources granted by HPC (π) from Shanghai Jiao Tong University.